\documentclass[bn,pdftex]{jjap3}
\usepackage{txfonts}
\usepackage{amsmath}
\usepackage{bm}
\usepackage{braket}
\usepackage{color}

\newcommand{\abs}[1]{\left|#1\right|}

\def\XXint#1#2#3{{\setbox0=\hbox{$#1{#2#3}{\int}$}
     \vcenter{\hbox{$#2#3$}}\kern-.5\wd0}}

\title{Analytic Expression for Magnetic Activation Energy}

\author{Daisuke Miura\thanks{dmiura@solid.apph.tohoku.ac.jp} and Akimasa Sakuma}
\inst{Department of Applied Physics, Tohoku University, Sendai 980-8579, Japan}

\abst{%
We theoretically investigate the magnetic activation energy of permanent magnets.
Practically, it is widely used in a phenomenological form as
$
\mathcal{F}_\mathrm{B}(H_\mathrm{ext})=\mathcal{F}_\mathrm{B}^0\left(1-H_\mathrm{ext}/H_0\right)^n,
$
where
$\mathcal{F}_\mathrm{B}^0$ is the activation energy in the absence of an external magnetic field $H_\mathrm{ext}$,
$n$ is a real parameter,
and $H_0$ is defined by the equation $\mathcal{F}_\mathrm{B}(H_0)=0$.
We derive the general and direct expressions for these phenomenological parameters
under the restriction of uniform rotation of magnetization
and
on the basis of the perturbative theory with respect to $H_\mathrm{ext}$.
Further,
we apply our results to Nd$_2$Fe$_{14}$B magnets
and confirm the validity of the proposed method by comparing with the Monte Carlo calculations.
}

\begin{document}
\maketitle

Elucidating the dominant factors in the coercive force of permanent magnets
is a central issue in the fields of magnetics and material science.
Magnetocrystalline anisotropy (MA) is one of the dominant factors governing the coercive force in rare-earth (RE) magnets such a Nd--Fe--B magnet,\cite{Sagawa1984,Croat1984}
whose temperature dependence has been investigated by many authors.\cite{Herbst1991,Skomski1999,Coey2010,Miyake2018}
From the theoretical viewpoint,
the MA of a ferromagnet is specified by its free energy density as a function of the magnetization angle,
and practically, its temperature dependence is expressed in terms of $\ell$th--order MA constants (MACs), $K_\ell(T)$, at a temperature $T$.
Especially, in RE magnets,
it is often found that those have higher order MACs and strongly depends on temperature.
Fortunately, these complex features can be understood within mean field theories (MFTs).\cite{Kuzmin2007,Sasaki2015,Miura2015,Ito2016,Miura2018,Miura2018c}
Most recently\cite{Miura2018},
we described the temperature-dependent MA in local moment systems by using Zener's phenomenological theory\cite{Zener1954}
and
derived it in an extended form of the Akulov--Zener--Callen--Callen power law,\cite{Akulov1936,Zener1954,Callen1966}
which is used to obtain a temperature dependence curve of $K_\ell(T)$ later in this study; there, it is referred to as the ``extended power law (EPL) .''
On the other hand,
it was reported that
inhomogeneity in magnetic structures seriously affects the coercive forces,\cite{Mitsumata2011}
and thus numerical analyses have been continued to date.\cite{Nishino2015,Nishino2018,Toga2016,Matsumoto2016,Nishino2017,Toga2018}

As mentioned above, the temperature dependence of MA in RE magnets
has been understood well.
However, the role of MA in the coercive force mechanism is not clear at this stage even within the MFT.
Especially, in the nonzero temperature range,
the coercive force depends on the observation time.
For this problem, Gaunt gave a direct answer by applying the Arrhenius formula\cite{Arrhenius1889}
to magnetization reversal dynamics,
in which the magnetization reversal time is characterized by a magnetic activation energy density.\cite{Gaunt1976,Gaunt1983,Gaunt1986}
Here, the activation energy density was proposed to have a form as\cite{El-Hilo2002}
\begin{align}
\mathcal{F}_\mathrm{B}(H_\mathrm{ext})=\mathcal{F}_\mathrm{B}^0(1-H_\mathrm{ext}/H_0)^n,
\label{eq:DF}
\end{align}
where
$H_\mathrm{ext}$ is the amplitude of the external magnetic field,
$\mathcal{F}_\mathrm{B}^0:=\mathcal{F}_\mathrm{B}(0)$,
$n$ is a real parameter,
and
$H_0$ is the amplitude of the external magnetic field required to cause magnetic reversal without the thermal activation.
$\mathcal{F}_\mathrm{B}(H_\mathrm{ext}) (>0)$
is defined by subtracting the initial value of the free energy density from the maximum value of one in a magnetization reversal process.
Therefore, $\mathcal{F}_\mathrm{B}(H_\mathrm{ext})$ depends on the path of the magnetization reversal process;
in other words,
$\mathcal{F}_\mathrm{B}^0, H_0$, and $n$ have the information about the magnetization reversal process,
and thus,
the investigation of these quantities is one of the good methods for understanding the coercive force mechanism from MA.
\cite{Givord1987,Givord1988,Victora1989,El-Hilo2002,Bance2015,Goto2015}

In the present study,
we aim to reveal the relation between the activation energy and MA
by explicitly representing $\mathcal{F}_\mathrm{B}^0, H_0$, and $n$
in terms of a given free-energy density,
in which we perform a perturbative calculation with respect to $H_\mathrm{ext}$.
Furthermore, we examine the validity of the perturbative result
by comparing it with the non-perturbative one obtained by the Monte Carlo (MC) methods.\cite{Toga2016}

First, we derive expressions for $\mathcal{F}_\mathrm{B}^0, H_0$, and $n$
in terms of the free energy density in a magnet.
In this study,
we assume that the free energy density in the absence of an external field is given
in the form of $F(\theta)$, which limits our discussion to the homogeneous magnetization-reversal process.
Taking the initial angle as $\theta=\theta_1$
and the most unstable angle in the process as $\theta=\theta_2$,
the angle-dependent free energy density satisfies
\begin{align}
F'(\theta_1)=F'(\theta_2)=0,\quad
F''(\theta_1)>0,\quad
F''(\theta_2)<0.
\end{align}
Now, applying the external magnetic field $H_\mathrm{ext}>0$ to the magnet,
the total free-energy density is given as $F(\theta)-\mu_0 M H_\mathrm{ext}\cos\theta$,
where $\theta$ is measured from the field, $\mu_0$ is the vacuum permeability, and $M$ is the saturation magnetization.
Here we notice that the extremal points depend on $H_\mathrm{ext}$ as $\theta_i\to\Theta_i(H_\mathrm{ext})$,
where $\Theta_i(0)\equiv\theta_i$.
Then, the magnetic activation energy density in the presence of $H_\mathrm{ext}$
is defined by
$
\tilde{\mathcal{F}}_\mathrm{B}(H_\mathrm{ext})
:=
F(\Theta_2)-F(\Theta_1)
-\mu_0 M H_\mathrm{ext}\left(
\cos \Theta_2
-
\cos \Theta_1
\right)
$, and perturbatively expanding $\tilde{\mathcal{F}}_\mathrm{B}(H_\mathrm{ext})$ with respect to $H_\mathrm{ext}$,
we obtain
\begin{align}
\tilde{\mathcal{F}}_\mathrm{B}(H_\mathrm{ext})
=
\tilde{\mathcal{F}}_\mathrm{B}(0)
\Biggl[
1-
\frac{\mu_0 M (\cos\theta_2-\cos\theta_1)}{\tilde{\mathcal{F}}_\mathrm{B}(0)}H_\mathrm{ext}
\notag\\
-
\frac{(\mu_0 M)^2}{2\tilde{\mathcal{F}}_\mathrm{B}(0)}
\left(
\frac{\sin^2\theta_2}{F''(\theta_2)}
-
\frac{\sin^2\theta_1}{F''(\theta_1)}
\right)
H_\mathrm{ext}^2
\Biggr]
+\mathcal{O}(H_\mathrm{ext}^3),
\label{eq:Fexpand}
\end{align}
where $\tilde{\mathcal{F}}_\mathrm{B}(0):=F(\theta_2)-F(\theta_1)$.
On the other hand,
the perturbative expansion for Eq. (\ref{eq:DF}) is given by
\begin{align}
\mathcal{F}_\mathrm{B}(H_\mathrm{ext})
=
\mathcal{F}_\mathrm{B}^0
\left(
1-\frac{n}{H_0}H_\mathrm{ext}
+
\frac{n(n-1)}{2H_0^2}H_\mathrm{ext}^2
\right)
+
\mathcal{O}(H_\mathrm{ext}^3).
\label{eq:Pexpand}
\end{align}
Therefore, the identity of $\tilde{\mathcal{F}}_\mathrm{B}(H_\mathrm{ext})\equiv \mathcal{F}_\mathrm{B}(H_\mathrm{ext})$
yields the following equations:
\begin{subequations}
\begin{align}
\mathcal{F}_\mathrm{B}^0&=F(\theta_2)-F(\theta_1),\\
n&=
\left[
1
+
\frac{\mathcal{F}_\mathrm{B}^0}{(\cos\theta_2-\cos\theta_1)^2}
\left(
\frac{\sin^2\theta_2}{F''(\theta_2)}
-
\frac{\sin^2\theta_1}{F''(\theta_1)}
\right)
\right]^{-1}
,\\
E_0&:=\mu_0 H_0 M
=
\frac{n\mathcal{F}_\mathrm{B}^0}{\cos\theta_2-\cos\theta_1}.
\end{align}
\label{eq:general}%
\end{subequations}
These are general expressions in the second-order $H_\mathrm{ext}$.
Here we notice that in the present theory,
there is no need to assume that the zero-field angles $\theta_i$ is small,
although the difference, $\Theta_i(H_\mathrm{ext})-\Theta_i(0)$, must be small.
Then, we can consider the usual case of $\theta_1=\pi$ and $\theta_2=\pi/2$,
and from Eqs. (\ref{eq:general}) one can obtain
\begin{subequations}
\begin{align}
\mathcal{F}_\mathrm{B}^0&=F(\pi/2)-F(\pi),\\
n&=
\left(
1
+
\frac{\mathcal{F}_\mathrm{B}^0}{F''(\pi/2)}
\right)^{-1}
,\\
E_0
&=
n\mathcal{F}_\mathrm{B}^0.
\end{align}
\label{eq:nosrt}%
\end{subequations}
Equations (\ref{eq:general}) and (\ref{eq:nosrt}) represent one of the main results in the present study.

Next, let us consider a practically important case of an angle-dependent free energy density given by
\begin{align}
F(\theta)=K_1\sin^2\theta+K_2\sin^4\theta.
\label{eq:F2}
\end{align}
From $F'(\theta)=0$,
it is possible to obtain a solution,
except for the trivial angles $\theta=0,\pi/2$, and $\pi$,
satisfying
\begin{align}
\sin^2\theta_\mathrm{SRT}=-\frac{K_1}{2K_2}.
\label{eq:srt}
\end{align}
When Eq. (\ref{eq:srt}) and $K_1\le 0$ are simultaneously satisfied,
$\theta=\theta_\mathrm{SRT}$ is a minimum angle;
hereinafter, this condition is referred to as the ``spin reorientation transition (SRT) condition''.
When the SRT condition is satisfied,
the most stable angle is given by $\theta=\theta_\mathrm{SRT}$,
that is, $\theta_1=\theta_\mathrm{SRT}$ and $\theta_2=\pi/2$, and accordingly, from Eq. (\ref{eq:general}) one can obtain
\begin{subequations}
\begin{align}
\mathcal{F}_\mathrm{B}^0&=K_2\left(1+\frac{K_1}{2K_2}\right)^2,
\\
n&=\frac{8}{5},\\
E_0&=\frac{8K_2}{5}\left(
1+
\frac{K_1}{2K_2}
\right)^{3/2}.
\end{align}
\label{eq:low}%
\end{subequations}
In the case of $K_1\ge 0$ and $K_2\ge 0$,
the SRT condition is not satisfied, and thus the result is that $\theta_1=\pi$ and $\theta_2=\pi/2$.
Then from Eqs. (\ref{eq:nosrt}), one can get
\begin{subequations}
\begin{align}
\mathcal{F}_\mathrm{B}^0&=K_1+K_2,
\\
n&=
2
\frac{
K_1+2K_2
}{K_1+3K_2}
,\\
E_0
&=
2K_1
\frac{
(1+2K_2/K_1)
(1+K_2/K_1)
}{
1+3K_2/K_1
}
.
\end{align}
\label{eq:high}%
\end{subequations}%
Here we notice that the well-known exact solution in the Stoner--Wohlfarth model\cite{Stoner1948} is reproduced by putting $K_2=0$.

\begin{figure}[tb]
\centering
\includegraphics[width=0.4\textwidth]{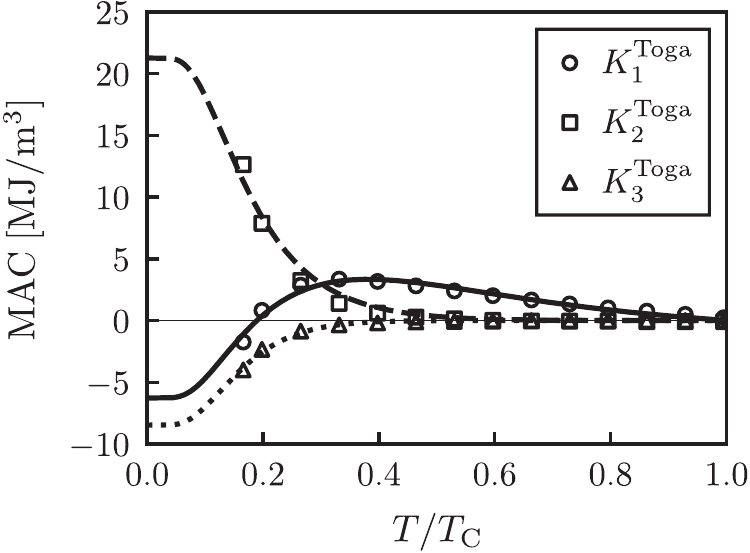}
\caption{EPL (lines) calculated by fitting to the $\ell$th-order MACs (open symbols) evaluated via the constrained MC method,\cite{Toga2016}
as a function of temperature.
The solid, dashed, and dotted lines denote $K_1^\mathrm{EPL}(T)$, $K_2^\mathrm{EPL}(T)$, and $K_3^\mathrm{EPL}(T)$, respectively.
}
\label{fig:K}
\end{figure}
\begin{figure}[tb]
\centering
\includegraphics[width=0.4\textwidth]{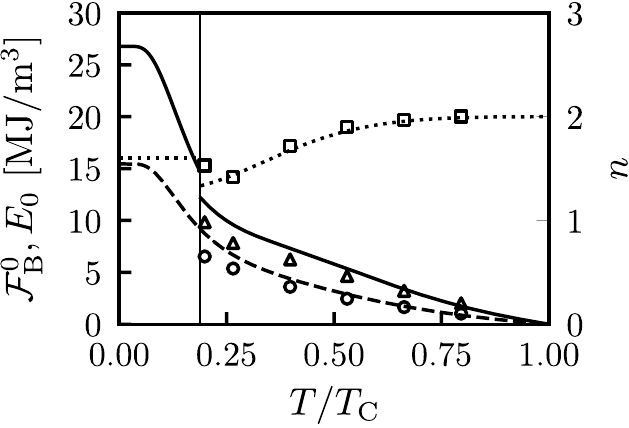}
\caption{Calculated phenomenological parameters for the magnetic activation energy density as a function of temperature.
The dashed, solid, and dotted lines denote $\mathcal{F}_\mathrm{B}^0, E_0$, and $n$, respectively.
The open symbols represent the nonpurturbative results evaluated by Toga et al.\cite{Toga2016},
in which the circles, triangles, and squares correspond to $\mathcal{F}_\mathrm{B}^0, E_0$, and $n$, respectively.
}
\label{fig:AE}
\end{figure}

Finally,
we examine the validity of our results
by comparing with the nonperturbative ones obtained with the MC method.
Toga et al.\cite{Toga2016}
evaluated the temperature-dependent MACs, $K_\ell^\mathrm{Toga}(T)$, up to the third order
by using the constrained classical MC method\cite{Asselin2010} in Nd$_2$Fe$_{14}$B magnets,
which is denoted by the open symbols in Fig. \ref{fig:K}.
To estimate the values of the MACs over the entire temperature range,
we can use the EPL as\cite{Miura2018}
\begin{subequations}
\begin{align}
K_1^\mathrm{EPL}(T)
&=
K_1^\mathrm{EPL}(0)\mu_\mathrm{Nd}(T)^3
\notag\\
&+
\frac{8}{7}
K_2^\mathrm{EPL}(0)
\left[
\mu_\mathrm{Nd}(T)^3
-
\mu_\mathrm{Nd}(T)^{10}
\right]
\notag\\
&+
\frac{8}{7}
K_3^\mathrm{EPL}(0)
\left[
\mu_\mathrm{Nd}(T)^3
-
\frac{18}{11}\mu_\mathrm{Nd}(T)^{10}
+
\frac{7}{11}\mu_\mathrm{Nd}(T)^{21}
\right],
\label{eq:EPL1}
\\
K_2^\mathrm{EPL}(T)
&=
K_2^\mathrm{EPL}(0)\mu_\mathrm{Nd}(T)^{10}
\notag\\
&+
\frac{18}{11}
K_3^\mathrm{EPL}(0)
\left[
\mu_\mathrm{Nd}(T)^{10}
-
\mu_\mathrm{Nd}(T)^{21}
\right],
\label{eq:EPL2}
\\
K_3^\mathrm{EPL}(T)
&=
K_3^\mathrm{EPL}(0)
\mu_\mathrm{Nd}(T)^{21},
\label{eq:EPL3}
\end{align}
\label{eq:EPL}%
\end{subequations}
where $\mu_\mathrm{Nd}(T)$ is given by
\begin{align}
\mu_\mathrm{Nd}(T)
&:=B_J\left(\frac{2\abs{g-1}J H_\mathrm{Nd}\mu_\mathrm{Fe}(T)}{T}\right),\\
B_J(x)&:=\frac{2J+1}{2J}\coth\left(\frac{2J+1}{2J}x\right)-\frac{1}{2J}\coth\left(\frac{x}{2J}\right),\\
\mu_\mathrm{Fe}(T)&:=\left[1-0.5\left(\frac{T}{T_\mathrm{C}}\right)^{3/2}-0.5\left(\frac{T}{T_\mathrm{C}}\right)^{5/2}\right]^{1/3},
\end{align}
and $J=9/2$, $g=8/11$, $H_\mathrm{Nd}=350$K,\cite{Yamada1988} $T_\mathrm{C}=586$K;\cite{Hirosawa1986,Sagawa1987}for detail, see Ref. \citenum{Miura2018}.
Each fitting parameter $K_\ell^\mathrm{EPL}(0)$ is determined as
$K_1^\mathrm{EPL}(0)=-6.28$MJ/m$^3$,
$K_2^\mathrm{EPL}(0)=21.27$MJ/m$^3$, and
$K_3^\mathrm{EPL}(0)=-8.48$MJ/m$^3$
by fitting Eqs. (\ref{eq:EPL}) to $K_\ell^\mathrm{Toga}$.
As shown in Fig. \ref{fig:K},
we can observe that $K_\ell^\mathrm{Toga}$ well obeys EPL, reflecting the fact that $K_\ell^\mathrm{Toga}$ was obtained in a local moment model.
Furthermore, the EPL describes a plateau in the low-temperature range
that does not appear within the classical theory, as remarked by Toga et al.,
and our extrapolated values $K_1^\mathrm{EPL}(0)$ and $K_2^\mathrm{EPL}(0)$
are consistent with the experimental values $-8.86$MJ/m$^3$ and $23.85$MJ/m$^3$, respectively.\cite{Durst1986}
Now we examine the validity of Eqs. (\ref{eq:low}) and (\ref{eq:high}) by using Eqs. (\ref{eq:EPL1}) and (\ref{eq:EPL2});
Fig. \ref{fig:AE} shows the comparison of our perturbative results (lines) with the nonperturbative ones (open symbols),
and we can observe that our present results consist with those obtained with the MC method by Toga et al.
One of the important conclusions in the presence of both $K_1$ and $K_2$ is
the point that the value of $n$ considerably deviates from Stoner-Wohlfarth's ``$2$'',
even in the simplest uniform rotation.
In addition,
our results demonstrate the presence of clear jumps in $n$ and $E_0$ at a temperature $T=T_\mathrm{SRT}\sim 0.2 T_\mathrm{C}$,
which corresponds to the SRT temperature;
$n\to 8/5$ and $E_0\to 8 K_2(T_\mathrm{SRT})/5$ for $T\nearrow T_\mathrm{SRT}$,
and $n\to 4/3$ and $E_0\to 4K_2(T_\mathrm{SRT})/3$ for $T\searrow T_\mathrm{SRT}$.
In our macroscopic viewpoints,
it is clear that this anomalous behavior originates from the complex temperature dependence of the MA as shown in Fig. \ref{fig:K}.
Unfortunately, the MC data of $n$ and $E_0$ did not be given in the temperature range below $T_\mathrm{SRT}$,
and therefore further investigations are desired to completely reveal the temperature dependence of the magnetic activation energy in realistic systems.

In summary,
we derived simple analytic expressions for the phenomenological parameters describing the magnetic activation energy
within the perturbative theory with respect to the external field,
and confirmed that our present results for Nd$_2$Fe$_{14}$B magnets consists with the nonperturbative ones obtained with the MC method.

\begin{acknowledgment}
We would like to thank
Dr. Y. Toga
for the useful discussions. %
This work was supported by
JSPS KAKENHI Grant
No. 16K06702,
No. 16H02390,
No. 16H04322,
and No. 17K14800,
in Japan.
\end{acknowledgment}

\bibliographystyle{jpsj}
\bibliography{library}

\end{document}